\documentclass[11pt,twoside]{article}
\usepackage{graphicx}
\usepackage{amsmath}
\usepackage{amsthm}
\usepackage{amssymb}
\usepackage{mdframed}
\usepackage{algorithm}
\usepackage{algpseudocode}
\usepackage{array}
\usepackage{threeparttable}
\usepackage{booktabs}

\newcommand{\confname}{From Formal Languages to Information Security}
\newcommand{\confauthors}{Short authors name}

\usepackage{fancyhdr}
\title{\textbf{On the Usage of Verifiable Credentials in Privacy-Preserving Federated Analytics}}

\author{
    \textbf{Andreea-Elena Dr\u{a}gnoiu}$^1$$^,$$^2$ \quad \textbf{Ruxandra F. Olimid}$^1$$^,$$^2$ \\
    \small{$^1$ Department of Computer Science, University of Bucharest, Bucharest, Romania} \\
    \small{$^2$ Research Institute of the University of Bucharest (ICUB), Bucharest, Romania} \\
    \small \texttt{andreea-elena.panait@drd.unibuc.ro ruxandra.olimid@fmi.unibuc.ro}
}

\date{}

\begin{document}

\maketitle

% --- Abstract ---
\textbf{Abstract.} 
Privacy-Preserving Federated Analytics enables multiple nodes to collaboratively derive statistical insights without exchanging raw data. However, ensuring node authenticity and data validation (avoiding concerns such as identity tracking, data linkage, or data leakage) remain fundamental challenges. This paper offers some insights into the feasibility of defining an authenticated, integrity-preserving data model by coupling FPPA with Verifiable Credentials defined within the Self-Sovereign Identity framework.
%Federated Privacy-Preserving Analytics (FPPA) enables decentralized entities to collaboratively derive statistical insights without exchanging raw data. However, verifying node authenticity and compliance without causing identity tracking or data linkage risks remains a fundamental challenge. This paper offers insights on the feasibility of building an integrity-preserving computing architecture by coupling FPPA with Virtual Credentials (VCs) within the Self-Sovereign Identity (SSI) framework.

% --- Sections ---

%-----------------------
\section{Introduction}
\label{sec:intro}

The growth of decentralized data generation and/or data ownership—driven by the expansion of the Internet of Things (IoT), mobile and edge computing, and cross-institutional collaborations—has increased demand for large-scale data analytics. Traditional centralized data analytics models require aggregating raw data into centralized repositories, which poses privacy risks and, in turn, regulatory liabilities (e.g., the General Data Protection Regulation (GDPR) \cite{gdpr2016regulation}, the California Consumer Privacy Act (CCPA) \cite{ccpa2018act}). To mitigate these risks, Privacy-Preserving Federated Analytics (PPFA) has emerged as a dominant paradigm. PPFA is a technique that enables analytics on data from multiple sources without exposing the raw data itself \cite{wang2025survey}.
%As a technique that enables analysis of data from distinct sources without exposing it, 
While maintaining data privacy, PPFA improves the overall utility of the data by integrating multiple sources of information in several ways—from simple statistics such as counts and averages to more advanced techniques such as machine learning. Examples of real-world use cases for PPFA include email analytics, cybersecurity collaboration platforms, and private bidding.

PPFA inherently preserves data privacy, which is critical because the underlying data come from multiple distinct sources and are often highly sensitive. But despite its advantages, PPFA's decentralized nature raises other fundamental challenges, including trust and identity management at scale. End nodes that own the data must be correctly identified and authenticated, while avoiding tracking or linkage that would defeat the purpose of federated privacy. At the same time, the data itself should be authenticated and integrity-protected to avoid poisoning attacks.

PPFA also has utility in personal data analytics, such as demographic analytics (e.g., age, gender, nationality, educational level) or group-based policies (e.g., group access rights or group performance capabilities). 
While reducing to digital identity attributes, Verifiable Credentials (VCs) within the Self-Sovereign Identity (SSI) framework \cite{dragnoiu2020survey,w3cvc2024} are a prominent model that gives users complete control over their identities, enabling context-dependent, sensitive disclosure while reducing reliance on third parties and mitigating privacy risks. In addition, VCs provide, by construction, integrity guarantees on the underlying attributes.

These aspects make VCs a reliable basis for PPFA. While Verifiable Presentations (VPs) are traditionally thought of as collections or aggregations of VCs belonging to a single holder, federated analytics on VCs (or, more generally, VPs\footnote{For simplicity of exposure, for the rest of the paper we will refer to VCs.}) could further enhance this by providing proofs on aggregated data from multiple holders while maintaining privacy within the group. Or, at its basics, one could use PPFA to privately extract aggregated information from collections of VCs without exposing the underlying data.

As a result, this work provides an introductory overview of applying PPFA within the SSI framework by applying data analytics directly on the VCs/VPs. 
While existing literature typically addresses either the computation layer (PPFA) or the trust layer (SSI/VCs), our work bridges this gap by evaluating a unified framework in which collections of distributed VCs serve as the immediate input to PPFA. References to joint PPFA and SSI exit, for example in the healthcare domain \cite{naeem2026decentralized}, but mostly restrict to SSI-based identification and access control rather than using SSI's VCs as a direct source of data for federated analytics.

The remainder of this paper is organized as follows: Section \ref{sec:related_work} gives the preliminaries, following three main directions: PPFA, SSI, and regulatory compliance. Section \ref{sec:system_model} presents the architectural model, a joint approach from SSI and PPFA. Section \ref{sec:limitations} discusses advantages, but also limitations and trade-offs. Section \ref{sec:conclusions} concludes.

% \textcolor{red}{TODO - delete}
% This is a section. 

% Keep the paper to 8 pages long. Use a single file \textit{paper.tex}. Do not use distinct files for distinct sections!

% Refer to the references as follows: \cite{atanasiu2007binary}. Add the reference strings in \textit{refs.bib}.

%-----------------------
\section{Preliminaries}
\label{sec:related_work}

%The convergence of decentralized identity frameworks and distributed computation represents a rapidly evolving research domain. 
%This section reviews the literature across three key pillars: privacy-preserving analytics, regulatory-driven data minimization, and verifiable decentralized identity.

This section reviews the three key pillars: PPFA, SSI, and related regulations. 

\paragraph{Privacy-Preserving Federated Analytics (PPFA).} Federated Analytics (FA) has emerged as the standard paradigm for deriving mathematical utility from diverse data sources without requiring central collection of the raw data \cite{wang2025survey}.
PPFA enhances privacy preservation by directly preventing raw data transmission. PPFA enforces various mechanisms, including Differential Privacy, i.e., adding errors/perturbations locally before transmitting the data, or rather complex cryptographic primitives such as Secure Multi-Party Computation (SMPC) or Homomorphic Encryption (HE) \cite{wang2025survey}.

To avoid confusion, we note that FA differs from Federated Learning (FL), as they serve different purposes: FA computes metrics from distributed data, while FL collaboratively trains Machine Learning (ML) models. While FL can serve as a tool for FA, they represent different concepts.

\paragraph{Self-Sovereign Identity (SSI).}
SSI is a digital identity model that gives the entity full control of its identity data, eliminating the need for centralized trusted parties. The SSI paradigm defines the trust relationship between three parties \cite{dragnoiu2020survey,w3cvc2024,richter2026self}: (1) \textit{the holder} - the entity (user) that owns and manages its identity through a digital wallet using VCs/VPs; (2) \textit{the issuer} - the authority that issues and signs a VC; (3) \textit{the verifier} - a service or entity that verifies/validates the authenticity of the VCs. 
To avoid direct interaction between the issuer and the verifier during the validation phase, a Verifiable Data Registry (VDR), typically a blockchain, anchors information such as Decentralized Identifiers (DIDs) and DID documents \cite{w3cvc2024,dragnoiu2020survey,dragnoiu2024towards}. The exposure of DID documents on the blockchain does not harm security, as DID documents do not contain sensitive information.
SSI enables full data ownership and enhances \textit{selective disclosure}, meaning an entity provides only the necessary information for a given scenario without exposing the rest (e.g., do not expose the exact age but prove legal age).
VCs (or, more generally, VPs) can be a viable source of information in PPFA because they contain data (attributes that characterize the holder), while enhancing privacy by construction (e.g., selective/minimal disclosure).

%Self-Sovereign Identity (SSI) frameworks and the W3C Verifiable Credentials (VC) standard \cite{w3cvc2024} decouple public identity from attribute proof, allowing for contextual data minimization. However, dynamic state tracking and cryptographic validation across multiple actors introduces systemic overhead. Drăgnoiu and Olimid \cite{dragnoiu2024towards} proposed an immutable identity management model anchored on the Arweave decentralized storage network. Their framework utilizes permanent, cost-efficient blockchain storage to maintain credential status registries without relying on centralized, trackable lookups. 

%Federated analytics is not federated learning!
%Federated analytics has emerged as the standard paradigm for deriving mathematical utility from decentralized data silos without requiring central collection. The foundational foundations of communication-efficient federated learning were established by McMahan et al. \cite{mcmahan2017communication}, demonstrating that model training could be distributed across edge nodes. To mitigate privacy leakage via inversion attacks on global models, subsequent works integrated Differential Privacy (DP) \cite{naseri2022local} and Secure Multi-Party Computation (SMPC). Comprehensive surveys by Kairouz et al. \cite{kairouz2021advances} highlight open challenges in node verification, showcasing a persistent vulnerability: the difficulty of verifying participant authenticity without exposing identifiable tracking metadata.

\paragraph{Regulatory Compliance.}
Modern computational frameworks must comply with applicable regulations, e.g., the European Union's \textit{General Data Protection Regulation} (GDPR) \cite{gdpr2016regulation}. SSI is now well established and standardized \cite{w3cvc2024}, and it is supported by the European SSI Infrastructure \cite{ebsi_vc}, thus facilitating legal compliance.

%Modern computational frameworks must adapt to strict legal mandates, primarily the European Union's \textit{General Data Protection Regulation} (GDPR) and the United States' \textit{California Consumer Privacy Act} (CCPA). Previous research indicates that traditional pseudonimized IDs do not prevent data linkage attacks under GDPR definitions. 

%The CHIST-ERA PATTERN project \cite{pattern2023d21} explicitly formalizes use cases—such as private bidding and cybersecurity threat intelligence exchange—where strict architectural isolation is required. Achieving compliance necessitates shifting from persistent host-identity certificates to ephemeral, zero-knowledge attribute verification prior to any federated round orchestration.

%-------------------------------
%-------------------------------
%-------------------------------
\section{Architectural Model}
\label{sec:system_model}

We further investigate the case where PPFA aggregates information across collections of VCs from multiple holders simultaneously and present an architectural model, depicted in Figure \ref{fig:figure1}. For simplicity, we follow the architectures in \cite{druagnoiu2025addressing} (Figure 2) for SSI and \cite{wang2025survey} (Figure 2) for PPFA as closely as possible. The joint architecture reveals the following roles:

%-----------------------
\begin{figure}[t!]
\centering
    \includegraphics[width=\textwidth]{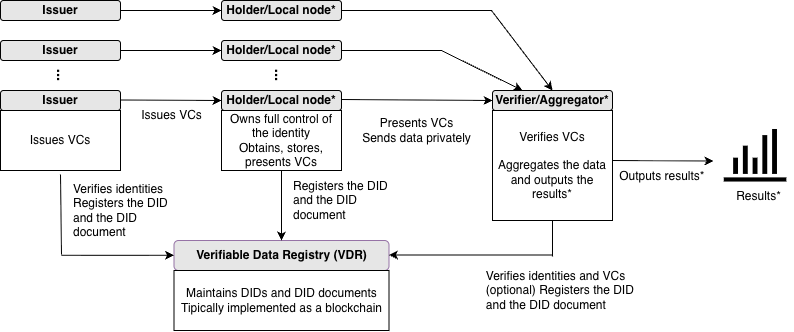}
\caption{SSI-based PPFA architectural model, following closely the SSI architecture from \cite{druagnoiu2025addressing} and the PPFA architecture from \cite{wang2025survey}. For easy identification, we mark by * the roles and operations originating from the PPFA architecture. }
\label{fig:figure1}
\centering
\end{figure}
%-----------------------

%Unlike traditional SSI setups, where a single holder interacts with a single verifier via a VP, we investigate the case where PPFA aggregates information across collections of VCs from multiple holders simultaneously.

%The proposed ecosystem transitions from classical node-identity matching to a dynamic multi-party trust framework. It consists of the following architectural roles:

\begin{itemize}
    \item \textbf{Issuer:} A trusted authority that signs and issues a VC to a holder after verifying their underlying claims or attributes. In our multi-node setting, one or more issuers may generate VCs for different holders.
    
    \item \textbf{Holder / Local node:} The entity that possesses one or more VCs. In our framework, following the line of PPFA, the holders are the local nodes that own the sensitive data; contrary to SSI, they do not present these credentials for authorization, but they act as the nodes that privately feed their local sensitive data, which is bound by the VCs, to the aggregator. By taking advantage of SSI capabilities, the nodes can enforce selective disclosure (i.e., provide only the minimal information needed for the required global computation) and cryptographically prove the validity of their data.
    %distributed processing units for the federated analytics pipeline, holding local sensitive data or attributes bound by the VCs.
    
    \item \textbf{Verifier / Aggregator:} The centralized or distributed orchestrator responsible for performing the analytics (e.g., computing averages, sums, or training ML iterations). %The aggregator establishes the baseline criteria for a given analytics round, 
    It validates incoming cryptographic proofs in the VCs without exposing individual attributes, confirming that the data originates from legitimate entities, is authentic, and has not been tampered with. Finally, it outputs the results.

    \item \textbf{Verifiable Data Registry (VDR):} The trust (decentralized) layer in SSI that allows the verifier to validate the data, without interacting with the issuer at the time of verification. It serves as a decentralized registry that hosts public keys, revocation registries, and schema definitions, and it is typically implemented as a blockchain to  provide general availability and immutability \cite{druagnoiu2025addressing,dragnoiu2024towards}.
    %Utilizing the immutable ledger paradigms established by Drăgnoiu and Olimid \cite{dragnoiu2024towards}, a permanent storage layer serves as a decentralized registry. This ledger hosts public keys, revocation registries, and schema definitions, eliminating single points of failure or trackable identity lookups.
\end{itemize}

%In addition, we present how the analytical workflow operates.

%\begin{enumerate}
%    \item \textbf{Query Initialization:} The aggregator broadcasts an analytics request alongside a specific attribute query parameter (e.g., ``Return proof of valid certification without revealing identity'').
%    \item \textbf{Local Computation and Proof Generation:} Each distributed holder processes the query locally against their sovereign Virtual Credential repository. Utilizing selective disclosure or zk-rpoofs, the node computes a local response and a validity proof. This proof guarantees the integrity of the underlying attribute and its active status on the decentralized ledger, adhering to the Drăgnoiu-Olimid identity model \cite{dragnoiu2024towards}.
%    \item \textbf{Confidential Aggregation:} The aggregator collects the response tuples and cryptographic proofs from participating nodes. By validating the proofs, the aggregator confirms that the incoming data originates from a legitimate, certified entity. The aggregator then executes the private federated analytics pipeline (e.g., computing a sum or an average) to generate the global analytics result without ever decoding the identity or tracking the history of any individual participant.
%\end{enumerate}

%-------------------------------
%-------------------------------
%-------------------------------

\section{Discussion}
\label{sec:limitations}

We recall that a joint view SSI-PPFA brings substantial architectural benefits. The cryptographic verifications inherited from the SSI framework ensure that the nodes' input is correct, authenticated, and integrity-protected. Verifiable certification by trusted authorities acts as a first layer of protection against malicious data injection or data poisoning.
By construction, the input data are (partially) protected, as the nodes can selectively disclose the information needed for the given federated computation. However, bringing these two distinct domains together introduces significant technical bottlenecks and opens up research challenges.

\paragraph{Computation and Communication Overhead.}
A primary limitation of the SSI-PPFA combined architecture is the cost it imposes on nodes, especially resource-constrained edge devices such as IoT nodes. Heavy cryptography, such as non-interactive zk-proofs and complex signatures and verifications, requires intensive CPU cycles and memory. By construction, the model relies on a VDR, typically a blockchain, which significantly increases complexity.

%One primary limitation of this combined architecture is the severe computational strain placed on resource-constrained edge devices, such as mobile phones or IoT nodes. Generating Non-Interactive Zero-Knowledge Proofs (NIZKPs) and managing pairing-based cryptographic operations require intense CPU cycles and memory allocation. If a federated analytics model requires hundreds of validation rounds, the local proof-generation delay can cause nodes to drop out of the cycle. Furthermore, while individual proofs are compact, broadcasting large verification payloads alongside model updates across thousands of nodes can saturate network bandwidth.

\paragraph{Portability and Framework Fragmentation.}
The SSI landscape is currently fragmented across competing cryptographic standards, decentralized ledger implementations, and DID methods \cite{druagnoiu2025addressing}. For a PPFA framework to run successfully across a diverse population of holders, it requires universal cross-platform portability. If different nodes hold VCs issued under incompatible cryptographic techniques or data models, the aggregator may struggle to compute results. Bridging this gap requires strict adherence to standardized data definitions, which remains an active industry hurdle, or using the proposed architecture in a smaller, specific homogeneous environment.

%The SSI landscape is currently fragmented across competing cryptographic standards, decentralized ledger implementations, and DID methods. For an PPFA framework to run successfully over a diverse population of holders, it requires universal cross-platform portability. If different nodes hold VCs issued under incompatible cryptographic curves or data models, the aggregator cannot initialize a uniform query. Bridging this gap requires strict adherence to standardized data definitions, which remains an active industry hurdle.

\paragraph{Dynamic Revocation Management at Scale.}
Maintaining anonymity while executing real-time credential revocation checks represents a major architectural trade-off. If an issuer revokes an employee's certification, the analytics framework must reflect that state change immediately to prevent problems such as data poisoning. However, looking up a credential's status in a traditional centralized registry lets the registry provider track when and where the credential is validated. While utilizing decentralized, permanent state registries like Arweave mitigates central tracking \cite{dragnoiu2024towards}, nodes must still download up-to-date cryptographic Accumulators or Cryptographic Revocation Lists (CRLs). Managing these large accumulator updates within rapid federated computation loops introduces a persistent trade-off between high privacy, correctness, and processing latency.

%Maintaining perfect anonymity while executing real-time credential revocation checks represents a major architectural trade-off. If an issuer revokes an employee’s certification, that state change must be immediately reflected in the analytics framework to prevent data poisoning or fraudulent bidding. However, looking up a credential's status in a traditional centralized registry allows the registry provider to track when and where the credential is being validated. While utilizing decentralized, permanent state registries like Arweave mitigates central tracking \cite{dragnoiu2024towards}, nodes must still download up-to-date cryptographic Accumulators or Cryptographic Revocation Lists (CRLs). Managing these large accumulator updates within rapid federated computation loops introduces a persistent trade-off between absolute privacy and processing latency.

\paragraph{Other Challenges.} A joint architecture faces risks and limitations from both SSI and PPFA. In depth, the blockchain's limitations and challenges, if used as a VDR, affect the overall system, bringing known problems such as scalability and interoperability. Finally, implementation challenges remain when considering detailed modeling or implementation. Our current work is incipient and only provides a high-level view of a joint SSI-PPFA model.

% This is a subsection. Do not use a single subsection in a section.

% Refer to Table \ref{tab:table1} in the text. Place all tables up [t!] or down [b!] on the page.

%-----------------------
% \begin{table}[b!]
% \centering
% \caption{Table caption}
% \medskip
% \label{tab:table1}
%     \begin{tabular}{lll}
%     \hline
%     \textbf{Column 1} & \textbf{Column2} & \textbf{Column 3}\\
%     \hline
%     R11 &  R12 & R13 \\
%     R21 &  R22 & R23 \\
%     \hline
%     \end{tabular}
% \end{table}
%-----------------------

%-----------------------
%\section{Discussions}

% This is another section. 

% Refer to Figure \ref{fig:figure1} in the text. Place all figures up [t!] or down [b!] on the page. 

% Place all figures in the \textit{images} folder!

% Colored images are not allowed!

\medskip

%-----------------------
\section{Conclusions}
\label{sec:conclusions}

This paper presented a foundational evaluation of integrating VCs within the SSI framework as a secure data layer for PPFA. The resulting joint SSI-PPFA model offers several advantages, including input-data validation and certified proofs, but also raises challenges such as higher costs, portability issues, and real-time synchronization of certified data.
Future work might include more detailed modeling and constructing a fully functional, open-source proof of concept integrated with decentralized storage networks.

%This paper presented a foundational evaluation of integrating Virtual Credentials (VCs) within the Self-Sovereign Identity (SSI) framework as a secure data layer for Federated Privacy-Preserving Analytics (FPPA).  Using the private decentralized bidding scenario aligned with the PATTERN project, we illustrated how collections of individual VCs can be leveraged to compute group-level statistics and verify compliance without leaking identity, specific certification parameters, or operational team structures. Future work will focus on constructing a fully functional, open-source proof of concept integrated with decentralized storage networks.

% --- Acknowledgements ---
\paragraph*{Acknowledgments.} This work was supported by a grant of the Ministry of Research, Innovation and Digitalization, CNCS/CCCDI - UEFISCDI, project number ERANET-CHISTERA-IV-PATTERN, within PNCDI IV.

% --- References ---
\bibliographystyle{splncs04}
\bibliography{refs}

\end{document}